\documentstyle[prl,aps,preprint,bezier]{revtex}
\begin{document}

\title{Renormalization group for multiparticle production in (2+1)
dimensions around the threshold}

\author{V. A. Rubakov and D. T. Son}

\address{Institute for Nuclear Research of the Russian Academy of Sciences,\\
         60th October Anniversary Prospect 7a, Moscow 117312 Russia.}
\date{July 1994}
\maketitle
\begin{abstract}
Multiparticle production in (2+1) dimensions is investigated. We show that
in a small region around the threshold the perturbation theory becomes
unapplicable due to infrared divergencies in a class of Feynman graphs with
rescattering in final state. We develop a technique based on
renormalization group for summing up leading logarithms and apply it to the
$\phi^4$ models with and without symmetry breaking and to the $O(N)$
$\phi^4$ theory.
\end{abstract}

\newpage

{\bf 1.} Multiparticle production in the conventional $\phi^4$ scalar
theory is attracting considerable interest. Though the initial obserbvation
\cite{Corn,Gold} was rather simple, only a few quantitative results has
been obtained so far. Most of them concern the multiparticle threshold at
which, for instance, formulas for the amplitude of production of $n$
particle by an initial virtual one has been derived at the tree
\cite{Vol1,Brown,AKP1} and one--loop levels \cite{Vol2,Smith,AKP2}. The
tree amplitude exhibits a factorial dependence on the number of created
bosons $n$. The correction from the first loop is of order $\lambda n^2$ in
comparison with the tree result (where $\lambda$ is the coupling constant)
and one can expect the breakdown of the perturbation theory at large $n$.
There have been several attempts to take into account contributions from
all orders of the perturbation theory \cite{GorVol,Mak}. In particular, the
amplitude has been calculated at leading order in $1/N$ in the $O(N)$
symmetric scalar theory at large N \cite{Mak}. However, it remains unclear
whether the $1/N$ expansion is applicable at large number of final
particles $n$.

In this paper we consider the multiparticle production in (2+1) dimensions,
where beside the above--mentioned inapplicability of the perturbation
theory at large $n$ there is another problem related with the infrared
divergence of a certain kind of loop graphs which breaks the conventional
pertubative expansion in a region close enough to the threshold. The graphs
of this kind contains loops related to the rescattering among final soft
particles. It is a pecular feature of (2+1) dimensions that right at the
threshold, rescattering graphs diverge logarithmically, so even at small
number of final particles $n$ calculation of the amplitude around the
threshold requires a nontrivial summing an infinite set of graphs.

We find a suitable technique to do such kind of summing which is a
modification of the conventional renormalization group (RG). This technique
allow us to sum up leading logarithms from all orders of the perturbation
theory. By this technique we obtain the amplitude $1\to n$ in the $\phi^4$
models with both broken and unbroken reflection symmetry, and in the $O(N)$
$\phi^4$ theory. We check the leading in $1/N$ amplitude in the latter
model and show explicitly that the $1/N$ expansion breaks down at large
number of final particles.

{\bf 2.} Let us consider the $\phi^4$ theory in (2+1) dimensions,
\[
   L = {1\over 2}(\partial_{\mu}\phi)^2 - {1\over 2}\phi^2
       -{\lambda\over 4}\phi^4,
\]
where, for convenience, we suppose that  the mass of the particle is equal
to unity. This is the theory of interacting relativistic bosons.

To describe bosons in the low--energy limit, one writes the following
effective lagrangian in term of a non--relativistic bosonic field $\Psi$,
\begin{equation}
  L_{\text{eff}}=
         \Psi^{\dagger}i\partial_0\Psi - {1\over 2}(\partial_i\Psi^{\dagger})
         (\partial_i\Psi) - g\Psi^{\dagger}\Psi^{\dagger}\Psi\Psi,
  \label{eff}
\end{equation}
where $g$ is some, yet to be determined, effective coupling

Non--relativistic bosons, interacting via a delta--like potential (as in
eq.\ (\ref{eff})), have been known for a long time as an example of a
non--relativistic system with dimensional transmutation \cite{Thorn}. In
fact, from dimension counting in (2+1)d (in non--relativistic kinematics
the relation between the dimensions of energy and momentum is
$[E]=[p]^{2}$) it is implied that $\Psi^{\dagger}\Psi^{\dagger}\Psi\Psi$ is
an marginal operator and the coupling constant $g$ is dimensionless.
Appearently, there is no scale parameter in the theory described by the
lagrangian (\ref{eff}). However, this scale exists and is simply the boson
mass (which is unity in our notation), which plays the role of the
``ultraviolet'' cutoff in the effective theory.

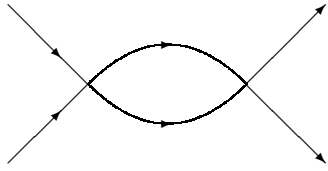
\begin{figure}
\caption{The only graph contributing to the lowest--order beta function
in the effective non--relativistic theory.}
\label{fig:beta}
\end{figure}

To make contact between the effective lagrangian and the initial
Lorentz--invariant one, one compares formulas for the amplitude of elastic
scattering of two bosons computed in both theories. This results the
following relation between $g$ and $\lambda$,
\begin{equation}
  g=3\lambda/8.
  \label{ini}
\end{equation}
Considering $g$ as the bare coupling, or the coupling at the scale of
ultraviolet cutoff, one can question about the evolution of $g$ as a
function of the momentum scale. For this purpose we introduce the running
coupling constant $g(t)$, which has the physical meaning of the strength of
the interaction between bosons at the momentum scale $p=\mbox{e}^{-t}$,
($t=\ln(1/p)$). There is only one Feymnan diagram (shown in fig.\
\ref{fig:beta}) that makes contribution to the corresponding beta function.
Simple calculations yield the following RG equation,
\[
  {dg(t)\over dt} = -{g^2(t)\over\pi},
\]
which has the solution
\begin{equation}
  g(t) = {g\over 1+ {g\over\pi}t} =
  {3\lambda\over 8}\left(1+{3\lambda\over 8\pi}\right)^{-1},
  \label{rgsol}
\end{equation}
where we have made use of the initial condition on $g(t)$, $g(0)=g$, and
$g$ is defined by eq.\ (\ref{ini}). From eq.\ (\ref{rgsol}) one sees that
the intensity of the interaction between bosons decreases as the momenta of
particles tend to zero. Later, we will demonstrate that this property holds
for a more general case of $O(N)$ $\phi^4$ model, but in the theory with
symmetry breaking the behavior of $g(t)$ is diametrically opposite.

The coupling constant changes considerably form its initial value only when
the momentum scale is exponentially small on $\lambda$, $t=\ln(1/p)\sim
1/\lambda$. So, the renormalization group is suitable for considering the
regime $\lambda\to 0$, $\lambda t\sim 1$. All further considerations will be
done for this particular regime.

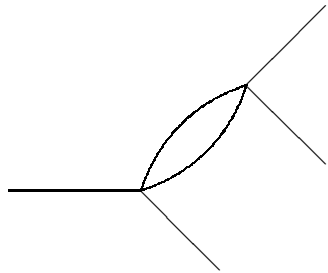
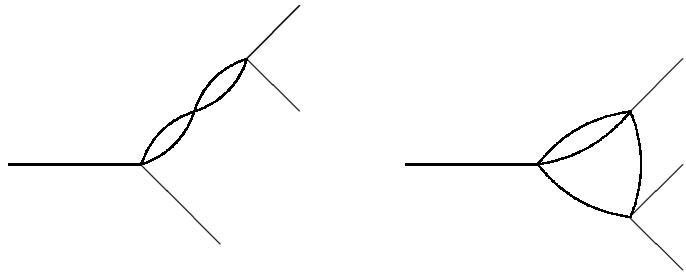
\begin{figure}
\caption{One-- and two--loops Feynman graphs that contribute to the
$1\to 3$ amplitude in the interesting regime.}
\label{fig:1to3}
\end{figure}

In fact the flow of the effective coupling, eq.\ (\ref{rgsol}), can be
obtained by a more trivial method by direct summing bubble graphs: in
non--relativistic theories only these graphs contribute to the elastic
scattering $2\to 2$. However, in calculation of the amplitude of
multiparticle production around the threshold more complicated diagrams are
involved and the problem cannot be reduced to the summing of bubble graphs.
For example, for the $1\to 3$ process the one-- and two--loop diagrams that
make contribution in our interesting regime is presented in fig.\
\ref{fig:1to3}. To deal with these processes the RG technique is required.
On the diagramatic language, the renormalization group corresponds to
summing leading logarithms, i.e. terms proportional to $(\lambda t)^n$ from
the whole perturbative series.

Let us try to describe the production of $n$ final particles in term of the
non--relativistic bosonic creation operator $\Psi^{\dagger}$. The only
relevant candidate is $A_n\Psi^{\dagger n}$. The following relation should
take place,
\[
  \langle n|\phi|0\rangle = \langle n|A_n\Psi^{\dagger n}|0
                             \rangle_{\text{eff}},
\]
where the matrix element in the l.h.s. is written in the initial $\phi^4$
model and the r.h.s. is understood as a matrix element in the effective
non--relativistic theory. Form this equation one finds that $A_n$ is equal
to the $1\to n$ amplitude when the spatial momenta of final particles is
small, but not exponentially small (so the logarithms do not appear in
loops). Let us for simplicity consider small enough $n$ where $A_n$
coincides with the tree $1\to n$ amplitude \cite{Vol1,Brown,AKP1},
\begin{equation}
  A_n = A_n^{\text{tree}} = n!\left({\lambda\over 8}\right)^{(n-1)/2}.
  \label{relation}
\end{equation}

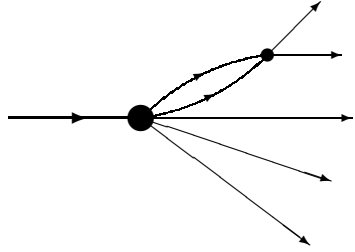
\begin{figure}
\caption{A graph that make contribution to RG equation for $1\to n$
amplitude. There are $n(n-1)/2$ diagrams of this type, corresponding to
different possiblities to choose of a pair of final particle to rescatter.}
\label{fig:1nrg}
\end{figure}

When characteristic scale of momenta of final particles is exponential, the
r.h.s. of eq. (\ref{relation}) is substantially renormalized by loops. One
can treat this renormalization by introducing the running vertex $A_n(t)$
and solving the corresponding RG equation,
\begin{equation}
  {dA_n(t)\over dt} = -{n(n-1)\over 2}{1\over\pi}g(t)A_n(t).
  \label{1neq}
\end{equation}
Graphically, this equation is represented in fig.\ \ref{fig:1nrg}.
Substituing the function $g(t)$ found in (\ref{rgsol}) to eq.\
(\ref{1neq}), the function $A(t)$ can be found easily,
\begin{equation}
  A_n(t) = A_n\left(1+{g(0)\over\pi}t\right)^{-{n(n-1)\over 2}}=
           A_n^{\text{tree}}\left(1+{3\lambda\over 8\pi}t\right)^
         {-{n(n-1)\over 2}}.
  \label{1nsol}
\end{equation}
So, we have found that the RG techniqe allows us to calculate the $1\to n$
amplitude $A_n(t)$ in a region close to the threshold where $\lambda t\sim
1$. Note that exactly at the threshold, i.e at $t=+\infty$, the amplitude
vanishes.

{\bf 3.} All the said above is equally applicable for the case of broken
symmetry,
\[
  L = {1\over 2}(\partial_{\mu}\phi)^2 -
      {\lambda\over 4}(\phi^2-v^2)^2,
\]
with the only exception that for calculating the bare coupling $g$ entering
into the lagrangian of the effective theory from the initial model one must
take into account not only the diagram with a four--boson vertex, but also
diagrams with two three--boson vertices, i.e. those with exchange of a
virtual particle in $s$--, $t$-- or $u$--channels. In contrast with the
case without symmetry breaking, the resulting amplitude is negative, which
means the attractive character of the force between boson at low energies.
One has
\[
  g = -{3\lambda\over 2}
\]
(we assume $v^2=(2\lambda)^{-1}$, so the mass of the boson is unity).
Formula (\ref{rgsol}) implies that the strength of the interaction
increases with $t$,
\begin{equation}
  g(t) = {g\over 1+ {g\over\pi}t} = -{3\lambda\over 2}
         \left(1-{3\lambda\over 2\pi}t\right)^{-1}.
  \label{growing}
\end{equation}
Litterally, eq.(\ref{growing}) predict an infinite coupling constant
at an exponentially small momentum scale of
\[
  p_0=\exp(-2\pi/3\lambda).
\]
This fact is a direct analog of the Landau pole in field theories without
asymtotic freedom. In reality, it is a manifestation of the existence of a
two--particle bound state in our model (recall that at least one bound
state exists in every, arbitrarily weak, two--dimensional attractive
potential). One can show that in our case, the energy of the bound state
is of order $p^2_0$.

The $1\to n$ amplitude is now
\begin{equation}
  A_n(t) = A_n^{\text{tree}}
              \left(1-{3\lambda\over 2\pi}t\right)
              ^{-{n(n-1)\over 2}}.
  \label{1nbroken}
\end{equation}
It is interesting to compare our result for the case of large $n$, $n\gg 1$
to that obtained in ref. \cite{GorVol} for (2+1) dimension by a different
method. In the case when $\lambda t$ is small, $\lambda t\ll 1$, our formula
in fact reproduces the result of \cite{GorVol}
\[
  A_n = A_n^{\text{tree}}\exp\left(
              {3\lambda\over 4\pi}n^2t\right).
\]
However, if $\lambda t$ is comparable with 1, the formula (\ref{1nbroken})
does not coincide with that obtained in ref.\ \cite{GorVol}. We consider it
as a counter--argument to the claim of ref.\ \cite{GorVol}.

{\bf 4.} Application of the technique described above the multi--component
$\phi^4$ model requires a slight modification. The lagrangian of the $O(N)$
$\phi^4$ model,
\begin{equation}
  L = {1\over 2}(\partial_{\mu}\phi_a)(\partial_{\mu}\phi_a)
      -{1\over 2}\phi_a\phi_a
      -{\lambda\over 4}(\phi_a\phi_a)^2,
  \label{ONlagr}
\end{equation}
($a=1\ldots N$ is the isospin index) contains one coupling constant,
$\lambda$. However, if one tries to write down the most general $O(N)$
symmetric effective non--relativistic lagrangian, one sees that there may
be two different effective couplings, $g_1$ and $g_2$, which correspond to
the two possible structures of the potential term,
\begin{equation}
  L_{\text{eff}} = \Psi_a^{\dagger}i\partial_0\Psi_a -
    {1\over 2}(\partial_i\Psi^{\dagger}_a)(\partial_i\Psi_a) -
    g_1\Psi_a^{\dagger}\Psi_a^{\dagger}\Psi_b\Psi_b -
    2g_2\Psi_a^{\dagger}\Psi_b^{\dagger}\Psi_a\Psi_b.
  \label{ONeff}
\end{equation}
$g_1$ is the low--energy elastic scattering scattering amplitude, singlet
in the $s$--channel, while $g_2$ determines amplitudes, singlet in $t$-- and
$u$--channels. From the initial lagrangian (\ref{ONlagr}) one obtains the
bare value of the coupling  constants,
\begin{equation}
  g_1 = g_2 = {\lambda\over 8}.
  \label{ONini}
\end{equation}
The fact that the bare values of $g_1$ and $g_2$ are equal is the remnant
of the crossing symmetry of our intial lagrangian (\ref{ONlagr}). However,
the evolution equations for the running coupling constants are rather
different,
\begin{equation}
  {dg_1(t)\over dt} = {1\over\pi}[Ng_1^2(t)+4g_1(t)g_2(t)],
  \label{ONrun1}
\end{equation}
\begin{equation}
  {dg_2(t)\over dt} = {2\over\pi}g_2^2(t).
  \label{ONrun2}
\end{equation}
In order to simplify the RG equations, let us introduce instead of $g_1(t)$
a linear combination of the two couplings,
\[
  g(t) = g_1(t) + {2\over N} g_2(t).
\]
The RG equation for $g(t)$ is simplier than that of $g_1(t)$,
eq.\ (\ref{ONrun1}),
\begin{equation}
  {dg(t)\over dt} = {N\over\pi}g^2(t).
  \label{ONrun3}
\end{equation}
The solution to the RG equations, eqs.\ (\ref{ONrun2}) and (\ref{ONrun3}),
which satisfies the initial condition (\ref{ONini}), can be found,
\[
  g_2(t) = {\lambda\over 8}\left(1+{\lambda\over 4\pi}t\right)^{-1},
\]
\[
  g(t) = \left(1+{2\over N}\right){\lambda\over 8}
         \left(1+(N+2){\lambda\over 8\pi}t\right).
\]

In analogy with the simple $\phi^4$ case, the production of $n$ soft bosons
from an initial particle with isospin $a$ can be described by an effective
operator $A_n\Psi_a^{\dagger} (\Psi_b^{\dagger}\Psi_b^{\dagger})^{(n-1)/2}$
($n$ must be odd). After some calculations we obtain the following RG
equation for $A_n(t)$,
\begin{equation}
  {dA_n(t)\over dt} ={1\over\pi}\left[
  {n-1\over 2}(N+n-1)g(t)+(n-1)^2\left(1-{1\over N}\right)g_2(t)\right]
  A_n(t).
  \label{ONrgeq}
\end{equation}
Having substitued the formulas for $g_2(t)$ and $g(t)$ to eq.\
(\ref{ONrgeq}) and solved it, one obtains the dependence of the $1\to n$
amplitude on the logarithm of the characteristic momentum of final
particles, $\ln(1/p)=t$
\begin{equation}
  A_n(t) = A_n^{\text{tree}}\left(1+(N+2){\lambda\over 8\pi}t\right)
           ^{-{n-1\over 2N}(N+n-1)}
           \left(1+{\lambda\over 4\pi}\right)
           ^{-{(n-1)^2\over 2N}(N-1)}.
  \label{ONfin}
\end{equation}
Recall that the regime where we are working is $\lambda\to 0$,
$\lambda t\sim 1$, provided other parameters as $n$ and $N$ are fixed.

Let us consider the large $N$ limit and show that the result of ref.\
\cite{Mak} can be reobtained in (2+1) dimensions by our technique.
Consider the case when $n$ much smaller than both the number of boson
flavors $N$ and the inverse coupling constant, eq.(\ref{ONfin}) reduces to
a much simplier formula,
\begin{equation}
  A_n(t) = A_n^{\text{tree}}\left[1+{N\lambda\over 8\pi}t\right]^{-(n-1)/2}.
  \label{largeN}
\end{equation}
Remember that the coupling constant enters into the tree amplitude by the
factor of $\lambda^{(n-1)/2}$, one can rewrite eq.(\ref{largeN}) into the
following form,
\begin{equation}
  A_n(t) \sim n! \lambda_R^{(n-1)/2},
  \label{makres}
\end{equation}
where
\[
  \lambda_R = {\lambda\over 1+{N\lambda\over 8\pi}t}
\]
is just the renormalized singlet scattering amplitude. One sees that for a
small number of final particles, $n\ll N$, leading order in $1/N$ result is
given just by the tree--level formula where the coupling constant is
replaced by the renormalized one. This is precisely the result of
\cite{Mak} in the particular case of (2+1) dimensions.

However, formula (\ref{makres}) has been deduced provided that $n\ll N$. If
the number of final particles is comparable with $N$, the effect of loops
is obviously not a simple renormalization of the coupling constant. One can
find from (\ref{ONfin}) that the correction to the leading order
(\ref{largeN}) is proportional to $n^2/N$. When the number of final
particles becomes comparable to the number of their sorts, the $1/N$
expansion becomes totally unreliable. One can expect that the breakdown of
the $1/N$ expansion is not a pecular feature of (2+1) dimensions but holds
also in the realistic (3+1) dimensional theory.

{\bf 6.} So, we see that the renormalization group is a poweful mean for
investigating the multiparticles amplitudes in (2+1) dimensional scalar
field theory at and around the threshold. The exact formula for the
amplitude, if ever be found, must incorporate the information obtained here
from the renormalization group equations.

\acknowledgements
The authors are indepted to Yu.~Makeenko, E.~Mottola, A.~Ovchinnikov,
L.~Yaffe for discussions of the results.

\newpage

\begin{picture}(120,80)(-150,0)
\put(0,20){\vector(1,1){20}}
\put(20,40){\line(1,1){10}}
\put(0,80){\vector(1,-1){20}}
\put(20,60){\line(1,-1){10}}
\bezier{200}(30,50)(60,80)(90,50)
\put(62,65){\vector(1,0){0}}
\bezier{200}(30,50)(60,20)(90,50)
\put(62,35){\vector(1,0){0}}
\put(90,50){\vector(1,-1){30}}
\put(90,50){\vector(1,1){30}}
\put(50,0){Fig.\ \ref{fig:beta}}
\end{picture}

\vskip 1.5in

\begin{picture}(130,120)(-150,0)
\thicklines
\put(0,60){\line(1,0){50}}
\thinlines
\bezier{100}(50,60)(60,90)(90,100)
\bezier{100}(50,60)(80,70)(90,100)
\put(50,60){\line(1,-1){30}}
\put(90,100){\line(1,1){30}}
\put(90,100){\line(1,-1){30}}
\end{picture}

\begin{picture}(260,120)(-90,0)
\thicklines
\put(0,60){\line(1,0){50}}
\thinlines
\bezier{100}(50,60)(55,75)(70,80)
\bezier{100}(50,60)(65,65)(70,80)
\bezier{100}(70,80)(75,95)(90,100)
\bezier{100}(70,80)(85,85)(90,100)
\put(90,100){\line(1,1){20}}
\put(90,100){\line(1,1){20}}
\put(90,100){\line(1,-1){20}}
\put(50,60){\line(1,-1){30}}

\thicklines
\put(150,60){\line(1,0){50}}
\thinlines
\bezier{100}(200,60)(213,77)(235,80)
\bezier{100}(200,60)(221,63)(235,80)
\bezier{100}(200,60)(213,43)(235,40)
\bezier{100}(235,80)(243,60)(235,40)
\put(235,80){\line(1,1){20}}
\put(235,40){\line(1,1){20}}
\put(235,40){\line(1,-1){20}}
\put(110,0){Fig.\ \ref{fig:1to3}}
\end{picture}

\newpage

\begin{picture}(120,130)(-160,0)
\thicklines
\put(0,80){\line(1,0){50}}
\put(30,80){\vector(1,0){0}}
\thinlines
\bezier{100}(50,80)(70,100)(98,104)
\bezier{100}(50,80)(78,84)(98,104)
\put(74,97){\vector(2,1){0}}
\put(78,89){\vector(2,1){0}}
\put(98,104){\vector(1,1){20}}
\put(98,104){\vector(1,0){28}}
\put(50,80){\vector(1,0){80}}
\put(50,80){\vector(3,-1){72}}
\put(50,80){\vector(4,-3){64}}
\put(50,80){\circle*{10}}
\put(98,104){\circle*{5}}
\put(40,0){Fig.\ \ref{fig:1nrg}}
\end{picture}

\end{document}